\documentclass[12pt,numbers]{article}

\usepackage{natbib}
\usepackage{amsmath}
\usepackage{amsthm}
\usepackage{amsfonts}
\usepackage{amssymb}
\usepackage{color}

\newlength{\xtrawidth}
\newlength{\xtraheight}
\newcommand{\Z}{\mathbb{Z}}
\DeclareMathOperator{\rank}{rank}

\newcommand{\ZZZ}{{\ensuremath{\Z_3\times\Z_3}}}
\newcommand{\Rep}[1]{\ensuremath{\mathbf{#1}}}
\newcommand{\barRep}[1]{\ensuremath{\overline{\Rep{#1}}}}

\newcommand{\dP}[1]{\ensuremath{dP_{#1}}}
\newcommand{\CP}[1]{\mathbb{P}^{#1}}
\newcommand{\B}[1]{\ensuremath{B_{#1}}}
\newcommand{\Xt}{{\ensuremath{\widetilde{X}}}}
\newcommand{\V}[1]{\ensuremath{{V}_{#1}}}
\newcommand{\Vt}{\ensuremath{\widetilde{V}}}
\newcommand{\W}[1]{{\ensuremath{{W}_{#1}}}}
\newcommand{\Osheaf}{\ensuremath{\mathcal{O}}}

\newcommand{\oB}[1]{\ensuremath{\Osheaf_{\B{#1}}}}
\newcommand{\oXt}{\ensuremath{\Osheaf_{\Xt}}}
\newcommand{\p}[1]{{\ensuremath{\pi_{#1}}}}

\begin{document}

\begin{titlepage}
  \begin{flushright}
    hep-th/0512177
    \\
    UPR-1141-T
  \end{flushright}
  \vspace*{\stretch{1}}
  \begin{center}
     \Huge 
     The Exact MSSM Spectrum\\ from String Theory
  \end{center}
  \vspace*{\stretch{2}}
  \begin{center}
    \begin{minipage}{\textwidth}
      \begin{center}
        \large         
        Volker Braun${}^{1,2}$, 
        Yang-Hui He${}^{3,4}$, 
        Burt A.~Ovrut${}^1$, 
        and Tony Pantev${}^2$
      \end{center}
    \end{minipage}
  \end{center}
  \begin{center}
    \begin{minipage}{\textwidth}
      \begin{center}
        ${}^1$ Department of Physics,
        ${}^2$ Department of Mathematics
        \\
        University of Pennsylvania,        
        Philadelphia, PA 19104--6395, USA
      \end{center}
      \begin{center}
        ${}^3$ Merton College, Oxford University,      
        Oxford OX1 4JD, U.K.
      \end{center}
      \begin{center}
        ${}^4$ Mathematical Institute, Oxford, 
        24-29 St.\ Giles', OX1 3LB, U.K.
      \end{center}
    \end{minipage}
  \end{center}
  \vspace*{\stretch{1}}
  \begin{abstract}
    \normalsize 
    We show the existence of realistic vacua in string theory whose
    observable sector has exactly the matter content of the MSSM. This
    is achieved by compactifying the $E_8 \times E_8$ heterotic
    superstring on a smooth Calabi-Yau threefold with an $SU(4)$ gauge
    instanton and a ${\mathbb Z}_{3} \times {\mathbb Z}_{3}$ Wilson
    line. Specifically, the observable sector is $N=1$ supersymmetric
    with gauge group $SU(3)_C \times SU(2)_L \times U(1)_Y \times
    U(1)_{B-L}$, three families of quarks and leptons, each family
    with a right-handed neutrino, and \emph{one} Higgs--Higgs
    conjugate pair.  Importantly, there are no extra vector-like pairs
    and no exotic matter in the zero mode spectrum. There are, in
    addition, $6$ geometric moduli and $13$ gauge instanton moduli in
    the observable sector. The holomorphic $SU(4)$ vector bundle of
    the observable sector is slope-stable.
  \end{abstract}
  \vspace*{\stretch{5}}
  \begin{minipage}{\textwidth}
    \underline{\hspace{5cm}}
    \centering
    \\
    Email: 
    \texttt{vbraun, ovrut@physics.upenn.edu},
    \texttt{yang-hui.he@merton.ox.ac.uk},
    \texttt{tpantev@math.upenn.edu}.
  \end{minipage}
\end{titlepage}

In a number of conference talks~\cite{talks}, we introduced a minimal
heterotic standard model whose observable sector has exactly the
matter spectrum of the MSSM. This was motivated and constructed as
follows.

The gauge group $Spin(10)$ is very compelling from the point of view
of grand unification and string theory since a complete family of
quarks and leptons plus a right-handed neutrino fits exactly into its
$\Rep{16}$ spin representation.  Non-vanishing neutrino masses
indicate that, in supersymmetric theories without exotic multiplets, a
right-handed neutrino must be added to each family of quarks and
leptons~\cite{neutrino}. Within the context~\cite{origin} of $N=1$
supersymmetric $E_8 \times E_8$ heterotic string vacua, 
a $Spin(10)$ group can arise from the
spontaneous breaking of the observable sector $E_8$ group by an
$SU(4)$ gauge instanton on an internal Calabi-Yau
threefold~\cite{WittenNew}. The $Spin(10)$ group is then broken by
discrete Wilson lines to a gauge group containing $SU(3)_C \times
SU(2)_L \times U(1)_Y$ as a factor~\cite{wilson}. To achieve this, the
Calabi-Yau manifold must have, minimally, a fundamental group $\ZZZ$.

Until recently, such vacua could not be constructed since Calabi-Yau
threefolds with fundamental group $\ZZZ$ and a method for building
appropriate $SU(4)$ gauge instantons on them were not known.  The
problem of finding elliptic Calabi-Yau threefolds with $\ZZZ$
fundamental group was rectified in~\cite{volker}. That of constructing
$SU(4)$ instantons was solved in a series of papers~\cite{hsm}, where
a class of $SU(4)$ gauge instantons on these Calabi-Yau manifolds was
presented. Generalizing the results in~\cite{extension,z2}, these
instantons were obtained as connections on rank $4$ holomorphic vector
bundles. In order for such connections to exist, it is necessary for
the corresponding bundles to be slope-stable. A number of non-trivial
checks of the stability of these bundles was presented in~\cite{hsm}.
A rigorous proof of the conjectured slope-stability recently appeared
in~\cite{stab}.  The complete low energy spectra were computed in this
context. The observable sectors were found to be almost that of the
minimal supersymmetric standard model (MSSM).  Specifically, the
matter content of the most economical of these vacua consisted of
three families of quarks/leptons, each family with a right-handed
neutrino, and \emph{two} Higgs--Higgs conjugate pairs.  Apart from
these, there were no other vector-like pairs, and no exotic particles.
That is, the observable sector is almost that of the MSSM, but
contains an extra pair of Higgs--Higgs conjugate fields.
Additionally, there are $6$ geometric moduli~\cite{volker} and $19$
vector bundle moduli~\cite{hsmm}. In~\cite{hsmmt}, it was shown that
non-vanishing $\mu$-terms can arise from cubic moduli-Higgs--Higgs
conjugate interactions. Despite the extra Higgs--Higgs conjugate
fields, the vacua presented in~\cite{hsm} are so close to realistic
particle physics that we refer to them as ``heterotic standard
models''.

These results were very encouraging. However, an obvious question is
whether one can, by refining these vector bundles, obtain
compactifications of the $E_{8} \times E_{8}$ heterotic string whose
matter content in the observable sector is \emph{exactly} that of the
MSSM.  The answer to this question is affirmative. In this paper, we
present models with an $N=1$ supersymmetric observable sector which,
for both the weakly and strongly coupled heterotic string, has the
following properties:

\subsection*{Observable Sector} 
\begin{itemize}
\item $SU(3)_C \times SU(2)_L \times U(1)_Y \times
  U(1)_{B-L}$ gauge group
\item Matter spectrum:
  \begin{itemize}
  \item $3$ families of quarks and leptons, each with a
    \emph{right-handed neutrino}
  \item $1$ Higgs--Higgs conjugate pair
  \item No exotic matter fields
  \item No vector-like pairs (apart from the one Higgs pair)
  \end{itemize}
\item $3$ complex structure, $3$ K\"ahler, and $13$ vector bundle moduli 
\end{itemize}
The holomorphic $SU(4)$ vector bundle $V$ leading to this observable sector
is slope-stable. A rigorous proof of this will be presented in~\cite{stab2}. 
Note that, although very similar to the supersymmetric standard model,
our observable sector differs in two significant ways. These are, first,
the appearance of an additional gauged $B-L$ symmetry and, second, the
existence of $6+13$ moduli fields, all uncharged under the gauge
group.
\bigskip

The structure of the hidden sector depends on the choice of a stable,
holomorphic vector bundle $V'$. The topology of $V'$, that is, its
second Chern class, is constrained by two conditions: first, the
anomaly cancellation equation
\begin{equation}
c_{2}\big(V'\big)=c_{2}\big(TX\big)-c_{2}\big(V\big)-[\mathcal{W}],
\label{bb1}
\end{equation}
where $[\mathcal{W}]$ is a possible effective five-brane class and, second,
a necessary condition of slope-stability given by
\begin{equation}
\int_{X}{ \omega \wedge c_{2}\big(V'\big)} >0
\label{bb2}
\end{equation}
for some K\"ahler class $\omega$. Often, this inequality is the only
obstruction to finding stable bundles. We expect that the second
condition is sufficiently strong that a subset of the bundles $V'$
satisfying it are slope-stable. Applying these conditions to the
specific Calabi-Yau threefold and $SU(4)$ observable sector bundle
discussed above, one can conclude the following.

\subsection*{Hidden Sector} 
\begin{itemize}
\item One expects there to exist holomorphic vector bundles $V'$ on
  the hidden sector which satisfy the anomaly cancellation condition
  and are slope-stable for K\"ahler classes $\omega$ for which the
  observable bundle $V$ is also stable.
\end{itemize}
We have not explicitly constructed such hidden sector bundles.  A
search for these is underway\footnote{Although exhibiting explicit
  $N=1$ supersymmetric hidden sectors is of interest, it is not clear
  that it is necessary, or even desirable, from the phenomenological
  point of view.  For example, supersymmetry breaking purely by
  gaugino condensation in the hidden sector may not lead to moduli
  stabilization with a small positive cosmological
  constant~\cite{Evgeny}. This might require the addition of
  anti-five-branes in the vacuum, as in~\cite{KKLT}, corresponding to
  an antieffective component of the five-brane class $[\mathcal{W}]$
  in the anomaly cancellation condition. Allowing anti-five-branes in
  the hidden sector would greatly simplify the search for stable
  hidden sector vector bundles.}. We will assume their existence in
the remainder of this paper.  \bigskip

The vacua presented above are a small subset of the heterotic standard
model vacua presented in~\cite{hsm}. As discussed below, their
construction involves subtleties in the analysis of the so-called
``ideal sheaf'' in the observable sector vector bundle, which were
previously overlooked. They appear to be the minimal such vacua, all
others containing either additional pairs of Higgs--Higgs conjugate
fields and/or vector-like pairs of families in the observable sector.
For this reason, we will refer to these vacua as ``minimal'' heterotic
standard models.

We note that, to our knowledge, these are the only vacua\footnote{At
  least until yesterday~\cite{ron}, when a nice generalization of the
  construction presented in~\cite{z2} (which makes stability manifest)
  appeared. Their model differs from ours in two respects. First, it
  uses a rank $5$ vector bundle instead of a rank $4$ one. Second,
  their one pair of Higgs fields arises in a codimension-two region in
  the moduli space, whereas our Higgs fields are generically present.}
whose spectrum in the observable sector has exactly the matter content
of the MSSM.  Other superstring
constructions~\cite{z2,chsw,greene,pheno-orb,pheno-II} lead to vacua
whose zero mode spectrum contains either exotic multiplets or
substantial numbers of vector-like pairs of Higgs and family fields,
or both. Although these might obtain an \emph{intermediate} scale mass
through cubic couplings with moduli (assuming these interactions
satisfy appropriate selection rules and the expectation values of the
moduli are sufficiently large), they can never be entirely removed
from the spectrum. To do so would violate the decoupling theorem.  For
these reasons, we speculate that heterotic standard models and, in
particular, the minimal heterotic standard model described in this
paper may be of phenomenological significance.

We now specify, in more detail, the properties of the these minimal
vacua and indicate how they are determined. The requisite Calabi-Yau
threefold, $X$, is constructed as follows~\cite{chsw}.  Let $\Xt$ be a
simply connected Calabi-Yau threefold which is an elliptic fibration
over a rational elliptic surface, $\dP9$. It was shown
in~\cite{volker} that $\Xt$ factors into the fiber product $\Xt=B_{1}
\times_{\mathbb{P}^1} B_{2}$, where $B_{1}$ and $B_{2}$ are both
$\dP9$ surfaces. Furthermore, $\Xt$ is elliptically fibered with
respect to each projection map $\pi_{i}:\Xt \rightarrow B_{i}$,
$i=1,2$. In a restricted region of their moduli space, such manifolds
can be shown to admit a $\ZZZ$ group action which is fixed-point free.
It follows that
\begin{equation}
  X=\frac{\Xt}{\ZZZ}
  \label{1}
\end{equation}
is a smooth Calabi-Yau threefold that is torus-fibered over a singular
$d\mathbb{P}_9$ and has non-trivial fundamental group
\begin{equation}
  \pi_{1}(X)=\ZZZ
  \,,
  \label{2}
\end{equation}
as desired. It was shown in~\cite{volker} that $X$ has 
\begin{equation}
  h^{1,1}(X)=3 \,, \quad 
  h^{2,1}(X)=3
  \label{3}
\end{equation}
K{\"a}hler and complex structure moduli respectively; that is, a total of $6$
geometric moduli.

We now construct a holomorphic vector bundle, $\V{}$, on $X$
with structure group
\begin{equation}
  G= SU(4)  
  \label{4}
\end{equation}
contained in the $E_8$ of the observable sector. For this bundle 
to admit a gauge connection satisfying the hermitian
Yang-Mills equations, it must be slope-stable. The connection 
spontaneously breaks the observable sector $E_8$ gauge symmetry to
\begin{equation}
  E_8 \longrightarrow Spin(10)
  \,,
  \label{5}
\end{equation}
as desired. We produce $\V{}$ by building stable, holomorphic vector bundles
$\Vt$ with structure group $SU(4)$ over $\Xt$ that are equivariant
under the action of $\ZZZ$. This is accomplished by generalizing the
method of ``bundle extensions'' introduced in~\cite{extension}. The
bundle $\V{}$ is then given as
\begin{equation}
  V=\frac\Vt{\ZZZ}
  \,.
  \label{6}
\end{equation}

Realistic particle physics phenomenology imposes additional
constraints on $\Vt$. Recall that with
respect to $SU(4) \times Spin(10)$ the adjoint representation of $E_8$
decomposes as
\begin{equation}
  \Rep{248}=
  \big( \Rep{1},\Rep{45} \big) \oplus  
  \big( \Rep{4},\Rep{16} \big) \oplus 
  \big( \barRep{4}, \barRep{16} \big) \oplus 
  \big( \Rep{6},\Rep{10} \big) \oplus
  \big( \Rep{15},\Rep{1} \big)  
  \,.
  \label{8}
\end{equation}
The number of $\Rep{45}$ multiplets is given by
\begin{equation}
  h^{0}\left(\Xt, \oXt \right)=1.
  \label{yes1}
\end{equation}
Hence, there are $Spin(10)$ gauge fields in the low energy theory, but
no adjoint Higgs multiplets. The chiral families of quarks/leptons
will descend from the excess of $\Rep{16}$ over $\barRep{16}$
representations. To ensure that there are three generations of quarks
and leptons after quotienting out $\ZZZ$, one must require that
\begin{equation}
  n_{\barRep{16}} -
  n_{\Rep{16}}
  =
  \frac{1}{2} c_{3}\big( \Vt \big)
  =
  -3 \cdot \big| \ZZZ \big| 
  = 
  -27
  \,,
  \label{7}
\end{equation}
where $n_{\barRep{16}}$, $n_{\Rep{16}}$ are the numbers of
$\barRep{16}$ and $\Rep{16}$ multiplets, respectively, and
$c_{3}(\Vt)$ is the third Chern class of $\Vt$.

The number of $\barRep{16}$ zero modes~\cite{z2} is given by
$h^1\big(\Xt,\Vt^*\big)$. Therefore, if we demand that there be no
vector-like matter fields arising from $\barRep{16}$-$\Rep{16}$ pairs,
$\Vt$ must be constrained so that
\begin{equation}
  h^1\left( \Xt, \Vt^* \right)=0
  \,.
  \label{9}
\end{equation}
Similarly, the number of $\Rep{10}$ zero modes is
$h^1\big(\Xt,{\wedge}^{2}\Vt\big)$. However, since the Higgs fields
arise from the decomposition of the $\Rep{10}$, one must not set the
associated cohomology to zero. Rather, we restrict $\Vt$ so that
$h^1\big(\Xt,{\wedge}^{2}\Vt\big)$ is minimal, but non-vanishing.
Subject to all the constraints that $\Vt$ must satisfy, we find that
the minimal number of $\Rep{10}$ representations is
\begin{equation}
  h^1\left( \Xt,{\wedge}^{2}\Vt \right)=4
  \,.
  \label{10}
\end{equation}
In~\cite{hsm}, the smallest dimension of this cohomology
group that we could find in the heterotic standard model context was
$h^1( \Xt,{\wedge}^{2}\Vt)=14$.  However, as discussed below, a more
detailed analysis of the ideal sheaf involved in the construction
of the vector bundle allows one to reduce this from $14$ to $4$.

We now present a stable vector bundle $\Vt$ satisfying
constraints eqns.~\eqref{7},~\eqref{9} and~\eqref{10}. 
This is constructed as an extension 
\begin{equation}
  0 
  \longrightarrow
  \V1
  \longrightarrow
  \Vt
  \longrightarrow
  \V2
  \longrightarrow
  0
  \label{C}
\end{equation}
of two rank $2$ bundles, $\V1$ and $\V2$. Each of these is the tensor
product of a line bundle with a rank $2$ bundle pulled back from a
$\dP9$ factor of $\Xt$. Using the two projection maps, we
define\footnote{See~\cite{hsm} for our notation of line bundles
  $\oXt(\cdots)$, etc.}
\begin{equation}
  \V1 =
  \oXt(-\tau_1+\tau_2) \otimes \p1^\ast(\W1)
  \,, \quad
  \V2 =
  \oXt(\tau_1-\tau_2) \otimes \p2^\ast(\W2)
  \,,
  \label{D}
\end{equation}
where
\begin{equation}
{\text span}\{\tau_{1},\tau_{2},\phi\}= H^{2}(\Xt,\mathbb{C})^{\ZZZ}
\label{burt1}
\end{equation}
is the $\ZZZ$ invariant part of the K\"ahler moduli space.
The two bundles, $\W1$ on $\B1$ and $\W2$ on $B_2$, are constructed
via an equivariant version of the Serre construction as
\begin{equation}
  0 
  \longrightarrow
  \chi_1
  \oB1(- f_1) 
  \longrightarrow
  \W1
  \longrightarrow
  \chi_1^2
  \oB1( f_1) \otimes I_3^{B_1}
  \longrightarrow
  0
  \
  \label{A}
\end{equation}
and
\begin{equation}
  0 
  \longrightarrow
  \chi_2^2
  \oB2(-f_2) 
  \longrightarrow
  \W2
  \longrightarrow
  \chi_2
  \oB2(f_2) \otimes I_6^{B_2}
  \longrightarrow
  0    
  \,,
  \label{B}
\end{equation}
where $I_3^{B_1}$ and $I_6^{B_2}$ denote the ideal sheaf\footnote{The
  analytic functions vanishing at the respective points.} of $3$ and
$6$ points in $B_1$ and $B_2$ respectively. Characters 
$\chi_1$ and $\chi_2$ are third roots of unity which generate the 
first and second factors of $\ZZZ$.

The crucial new observation occurs in the definitions of $W_{1}$ and
$W_{2}$. Satisfying condition eq.~\eqref{7} requires that one use
ideal sheaves of $9$ points in total. In our previous
papers~\cite{hsm}, we chose $W_{1}$ to be the trivial bundle and
defined $W_{2}$ as an extension of two rank $1$ bundles, one of which
contained a single ideal sheaf, $I_{9}$.  This comprises $9$ points,
as it must. However, it is possible to use several such sheaves in the
definitions of $W_{1}$ and $W_{2}$, as long as the total number of
points is $9$. Note that while the $\ZZZ$ action on $\Xt$ only has
orbits consisting of $9$ points, the $\ZZZ$ action on the base
surfaces $B_1$ and $B_2$ is not free and, in fact, has orbits of $9$
and of $3$ points. This allows one to define the ideal sheaf
$I_3^{B_1}$ using the fixed points of the second $\Z_3$ on $B_1$ and
the ideal sheaf $I_6^{B_2}$ using the fixed points of the second
$\Z_3$ on $B_2$ taken with multiplicity $2$. That is, previously we
only considered the case where the total of $9$ points were
distributed as\footnote{The ideal sheaf of $0$ points is just the
  trivial line bundle.} $0+9$. In this paper, we distribute the points
into two different ideal sheaves as $3+6$.  This allows us to obtain
the precise MSSM matter content.

We now extend the observable sector bundle $\V{}$ by adding a Wilson
line, $W$, with holonomy
\begin{equation}
  \mathrm{Hol}(W)=\ZZZ \subset Spin(10)
  \,.
  \label{12}
\end{equation}
The associated gauge connection spontaneously breaks $Spin(10)$ as
\begin{equation}
  Spin(10) \longrightarrow 
  SU(3)_{C} \times 
  SU(2)_{L} \times 
  U(1)_{Y} \times 
  U(1)_{B-L}
  \,,
  \label{13}
\end{equation}
where $SU(3)_{C} \times SU(2)_{L} \times U(1)_{Y}$ is the standard model gauge
group. Since $\ZZZ$ is Abelian and 
$\rank\big(Spin(10)\big)=5$, an additional rank one factor must appear.
For the chosen embedding of $\ZZZ$, 
this is precisely the gauged $B-L$ symmetry.

As discussed in~\cite{z2}, the zero mode spectrum of $\V{} \oplus W$
on $X$ is determined as follows. Let $R$ be a representation of
$Spin(10)$, and denote the associated $\Vt$ bundle by $U_{R}(\Vt)$.
Find the representation of $\ZZZ$ on $H^1\big(
\Xt,U_{R}(\Vt)\big)$ and tensor this with the representation of the
Wilson line on $R$. The zero mode spectrum is then the invariant
subspace under this joint group action. Let us apply this to the case
at hand.  To begin with, the single $\Rep{45}$ decomposes into the
$SU(3)_{C} \times SU(2)_{L} \times U(1)_{Y} \times U(1)_{B-L}$ gauge
fields. Now consider the $\barRep{16}$ representation. It follows from
eq.~\eqref{9} that no such representations occur. Hence, no $SU(3)_{C}
\times SU(2)_{L} \times U(1)_{Y} \times U(1)_{B-L}$ fields arising
from vector-like $\barRep{16}$-$\Rep{16}$ pairs appear in the
spectrum, as desired. Next examine the $\Rep{16}$ representation. The
constraints~\eqref{7} and~\eqref{9} imply that
\begin{equation}
  h^1\left( \Xt,\Vt \right)=27
  \,.
  \label{14}
\end{equation}
One can calculate the $\ZZZ$ representation on
$H^1\big(\Xt,\Vt\big)$, as well as the Wilson line action on
$\Rep{16}$. We find that
\begin{equation}
H^1\big(\Xt,\Vt\big)=RG^{\oplus3},
\label{burt2}
\end{equation}
where $RG$ is the regular representation of $G=\ZZZ$ given by
\begin{equation}
RG=1 \oplus \chi_{1} \oplus \chi_{2} \oplus \chi_{1}^{2} \oplus \chi_{2}^{2}
\oplus \chi_{1}\chi_{2} \oplus \chi_{1}^{2}\chi_{2} \oplus 
\chi_{1}\chi_{2}^{2} \oplus \chi_{1}^{2}\chi_{2}^{2}.
\label{burt3}
\end{equation}
Furthermore, the Wilson line action can be chosen so that
\begin{multline}
  \label{burt4}
  \Rep{16}= \Big[
  \chi_{1}\chi_{2}^{2} \big(\Rep{3}, \Rep{2}, 1, 1 \big)
  \oplus \chi_{2}^{2} \big(\Rep{1},\Rep{1}, 6, 3 \big)
  \oplus \chi_{1}^{2}\chi_{2}^{2} \big(\barRep{3},\Rep{1}, -4, -1 \big)\Big]
  \oplus \\ \oplus 
  \Big[  \big(\Rep{1},\Rep{2}, -3, -3 \big) 
  \oplus \chi_{1}^{2} \big(\barRep{3},\Rep{1}, 2, -1 \big)\Big]
  \oplus \chi_{2} \big(\Rep{1},\Rep{1}, 0, 3 \big).
\end{multline}
Tensoring these together, we find that the invariant
subspace consists of three families of quarks and leptons, each family
transforming as
\begin{equation}
  \big(\Rep{3},   \Rep{2}, 1, 1 \big) \,,\quad
  \big(\barRep{3},\Rep{1}, -4, -1 \big) \,,\quad
  \big(\barRep{3},\Rep{1}, 2, -1 \big)
  \label{15}
\end{equation}
and
\begin{equation}
  \big(\Rep{1},\Rep{2}, -3, -3 \big) \,,\quad
  \big(\Rep{1},\Rep{1}, 6, 3 \big) \,,\quad
  \big(\Rep{1},\Rep{1}, 0, 3 \big)
  \label{16}
\end{equation}
under $SU(3)_{C} \times SU(2)_{L} \times U(1)_{Y} \times U(1)_{B-L}$.
We have displayed the quantum numbers $3Y$ and $3(B-L)$ for
convenience. Note from eq.~\eqref{16} that each family contains a
right-handed neutrino, as desired.

Next, consider the $\Rep{10}$ representation. Recall from
eq.~\eqref{10} that $h^1\big(\Xt,{\wedge}^{2}\Vt\big)=4$. We find
that the representation of $\ZZZ$
in $H^1\big(\Xt,{\wedge}^{2}\Vt\big)$ is given by  
\begin{equation}
  H^1\big(\Xt,{\wedge}^{2}\Vt\big)= 
  \chi_2 \oplus \chi_2^2 \oplus 
  \chi_{1}\chi_{2}^{2} \oplus \chi_{1}^{2}\chi_{2}
  \,.
  \label{17}
\end{equation}
Furthermore, the Wilson line $W$ action is
\begin{equation}
  \Rep{10}= 
  \Big[
  \chi_2^2\big(\Rep{1},\Rep{2},3,0\big) \oplus 
  \chi_1^2\chi_2^2\big(\Rep{3},\Rep{1},-2,-2\big)
  \Big]
  \oplus 
  \Big[
  \chi_2\big(\Rep{1},\barRep{2},-3,0\big) \oplus 
  \chi_1\chi_2\big(\barRep{3},\Rep{1},2,2\big)
  \Big]
  \,.
  \label{19}
\end{equation}
Tensoring these actions together, one finds that the invariant
subspace consists of a single copy of
\begin{equation}
  \big( \Rep{1},\Rep{2}, 3, 0 \big) \,,\quad
  \big( \Rep{1},\barRep{2}, -3,  0 \big)
  \,.
  \label{21}
\end{equation}
That is, there is precisely one pair of Higgs--Higgs conjugate fields
occurring as zero modes of our vacuum.

Finally, consider the $\Rep{1}$ representation of the $Spin(10)$ gauge
group. It follows from~\eqref{8}, the above discussion, and the fact
that the Wilson line action on $\Rep{1}$ is trivial that the number of
$\Rep{1}$ zero modes is given by the $\ZZZ$ invariant subspace of
$H^1\big(\Xt, \Vt \otimes \Vt^* \big)$, which is denoted by
$H^1\big(\Xt, \Vt \otimes \Vt^* \big)^\ZZZ$. Using the formalism
developed in~\cite{hsmm}, we find that
\begin{equation}
  h^1\left(\Xt, \Vt \otimes \Vt^* \right)^\ZZZ=13.
  \label{moduli}
\end{equation}
That is, there are $13$ vector bundle moduli.

Putting these results together, we conclude that the zero mode
spectrum of the observable sector has gauge group $SU(3)_C \times
SU(2)_L \times U(1)_Y \times U(1)_{B-L}$, contains three families of
quarks and leptons each with a right-handed neutrino, has one
Higgs--Higgs conjugate pair, and contains no exotic fields or
additional vector-like pairs of multiplets of any kind.  Additionally,
there are $13$ vector bundle moduli.

As a final step, one must demonstrate that $\Vt$ is slope-stable. This
has been proven, in detail, and will be presented in~\cite{stab2}.
Here, suffice it to say that $\Vt$ will be stable with respect to any
K\"ahler class in a finite three-dimensional region of K\"ahler moduli
space containing the point
\begin{equation}
\omega= 3\big( 2 \tau_{1} + 3 \tau_{2} + \phi \big)
.
\label{burt5}
\end{equation}
Henceforth, we restrict our discussion to this region of moduli space,
which we denote by $\mathcal{K}^s$.

Another important constraint for realistic compactifications is the
existence of Yukawa couplings. Recall that (via the Kaluza-Klein
reduction) the massless fields are associated with a number of
vector-bundle valued harmonic one-forms $\Psi_i$ on the Calabi-Yau
threefold. Their Yukawa coupling is then given by the integral
\begin{equation}
  \label{eq:yukawa}
  \lambda_{ijk} = \frac{1}{9}\int_{\Xt} 
  \Omega \wedge
  \mathrm{Tr}
  \Big(
  \Psi_i \wedge \Psi_j \wedge \Psi_k 
  \Big)
  \,,
\end{equation}
where the $\mathrm{Tr}$ denotes a suitable contraction of the vector
bundle indices. The integral is only non-zero if the legs of the three
one-forms $\Psi_i$ span the $\pi_1$-fiber direction, the $\pi_2$ fiber
direction, and the base $\CP1$ direction. This is the case here. A
detailed analysis reveals that we do, indeed, have non-vanishing
Yukawa couplings~\cite{yukawa}.

Thus far, we have discussed the vector bundle of the observable
sector. However, the vacuum can contain a stable, holomorphic vector
bundle, $\Vt'$, on $X$ whose structure group is in the $E_8'$ of the
hidden sector. The requirement of anomaly cancellation relates the
observable and hidden sector bundles, imposing the constraint that
\begin{equation} 
  c_2\big( \Vt' \big)=
  c_2\big( T\Xt \big) -
  c_2\big( \Vt  \big) -
  [\mathcal{W}],
  \label{22}
\end{equation}
where $[\mathcal{W}]$ must be an effective class and $c_2$ is the 
second Chern class. In
the strongly coupled heterotic string, $[\mathcal{W}]$ is the class of
the holomorphic curve around which a bulk space five-brane is wrapped.
In the weakly coupled case $[\mathcal{W}]$ must vanish. We have
previously constructed $\Xt$ and $\Vt$ and, hence, can compute
$c_{2}\big(T\Xt\big)$ and $c_{2}\big(\Vt\big)$. They are found to be
\begin{equation}
c_{2}\big(T\Xt\big)=12 \big(\tau_{1}^{2} + \tau_{2}^{2} \big), \quad
c_{2}\big(\Vt\big)= \tau_{1}^{2} + 4\tau_{2}^{2} +4\tau_{1}\tau_{2}
\label{burt6}
\end{equation}
respectively. Inserting these results, eq.~\eqref{22}
becomes a constraint on the hidden sector bundle $\Vt'$. Henceforth, we assume
that $\Vt'$ satisfies~\eqref{22}. The easiest
possibility is that $\Vt'$ is the trivial bundle. However, in this
case, we find that $[\mathcal{W}]$ is not effective. Hence, we must choose
the hidden sector bundle $\Vt'$ to be non-trivial.

However, simply satisfying~\eqref{22} is not sufficient. As discussed
previously, $\Vt'$ must also be slope-stable. As a guide to
constructing stable, holomorphic vector bundles $\Vt'$ in the hidden
sector, we note the following condition. It can be shown that if
$\Vt'$ is slope-stable with respect to a K\"ahler class $\omega$, it
must satisfy the ``Bogomolov inequality''
\begin{equation}
\int_{\Xt}{c_{2}\big(\Vt'\big)\wedge \omega}> 0.
\label{burt7}
\end{equation}
Note that if $c_{2}\big(\Vt'\big)$ is Poincare dual to an effective
(antieffective) curve, then~\eqref{burt7} is satisfied (never
satisfied) for any choice of K\"ahler class. Most vector bundles
$\Vt'$ have a second Chern class whose Poincare dual is neither
effective nor antieffective. In this case, constraint~\eqref{burt7}
is satisfied for $\omega$'s contained in a non-vanishing subspace of
the K\"ahler cone. One can explicitly analyze this subspace using the
second Chern class derived from anomaly condition~\eqref{22}.  It is
simplest to limit our discussion to $\Vt'$ for which
$[\mathcal{W}]=0$.  The generalization to the case where
$[\mathcal{W}]$ is non-vanishing is straightforward. In this case,
eqns.~\eqref{22} and~\eqref{burt6} imply that
\begin{equation}
c_{2}\big(\Vt'\big)=11\tau_{1}^{2} + 8\tau_{2}^{2} - 4\tau_{1}\tau_{2}.
\label{burt8}
\end{equation}
Recalling from~\eqref{burt1} that $\tau_{1}$,$\tau_{2}$ and $\phi$ are
a basis for the $\ZZZ$ invariant K\"ahler moduli space, we can
parameterize an arbitrary such K\"ahler class by
\begin{equation}
\omega=x_{1}\tau_{1}+x_{2}\tau_{2}+y\phi.
\label{burt9}
\end{equation}
Then, using the relations $\tau_{1}^{3}=\tau_{2}^{3}=\phi^{2}=0$,
$\tau_{1}\phi=3\tau_{1}^{2}$ and $\tau_{2}\phi=3\tau_{2}^{2}$ we see 
using~\eqref{burt8} and~\eqref{burt9} that
\begin{equation}
c_{2}\big(\Vt'\big)\wedge \omega = 4x_{1}+7x_{2}-12y.
\label{burt10}
\end{equation}
It follows that constraint~\eqref{burt7} will be satisfied if
\begin{equation}
 4x_{1}+7x_{2}-12y > 0.
\label{burt11}
\end{equation}
This defines a three-dimensional region of moduli space which we
denote by $\mathcal{K}^B$. Note that the K\"ahler class~\eqref{burt5} for
which the observable sector bundle $\Vt$ was proven to be stable also
satisfies~\eqref{burt11}. Hence,
\begin{equation}
\mathcal{K}^s \cap \mathcal{K}^B \neq \emptyset.
\label{burt12}
\end{equation}
In fact, one can show that $\mathcal{K}^s \cap \mathcal{K}^B$ is a finite
three-dimensional subcone of the K\"ahler cone. It follows that both
$\Vt$ and $\Vt'$ can, in principle, be slope-stable with respect to
any K\"ahler class $\omega \in \mathcal{K}^s \cap \mathcal{K}^B$.

Fix $\omega \in \mathcal{K}^s \cap \mathcal{K}^B$. There are numerous
vector bundles $\Vt'$ with second Chern class~\eqref{burt8} which
satisfy condition~\eqref{burt7} for this choice of $\omega$.
Since~\eqref{burt7} is only a necessary condition for stability, we
expect that many such $\Vt'$ are not stable. Indeed, one can construct
explicit examples for which this is the case. However,~\eqref{burt7}
is a very strong condition and it is believed that at least some
$\Vt'$ are slope-stable with respect to $\omega$. Furthermore, since
one may choose any $\omega$ in the three-dimensional space
$\mathcal{K}^s \cap \mathcal{K}^B$, it becomes even more probable that
there exist slope-stable vector bundles $\Vt'$ with respect to at
least one such $\omega$.

We conclude that one expects that there should exist hidden sector
holomorphic vector bundles $\Vt'$ that satisfy the anomaly
cancellation condition and are slope-stable. Explicit examples of such
bundles will be presented elsewhere.


\paragraph{Acknowledgments}

We are grateful to E.~Buchbinder, R.~Donagi, P.~Langacker, B.~Nelson
and D.~Waldram for enlightening discussions. This research was
supported in part by cooperative research agreement DE-FG02-95ER40893
with the U.~S.~Department of Energy and an NSF Focused Research Grant
DMS0139799 for ``The Geometry of Superstrings''.  Tony Pantev is
partially supported by NSF grants DMS 0104354 and DMS 0403884.  Yang
Hui-He is supported in part by the FitzJames Fellowship at Merton
College, Oxford.


\end{document}